\documentclass[12pt]{article}
\usepackage{amsmath}
\usepackage{amssymb}

\newcommand{\be}{\begin{equation}}
\newcommand{\bea}{\begin{eqnarray}}
\newcommand{\ee}{\end{equation}}
\newcommand{\eea}{\end{eqnarray}}

\def\theequation{\arabic{section}.\arabic{equation}}
\newcommand{\bn}[2]{\left(\begin{array}{c} #1\\ #2
\end{array}\right)}

\begin{document}
\topmargin -1cm \oddsidemargin=0.25cm\evensidemargin=0.25cm
\setcounter{page}0
\renewcommand{\thefootnote}{\fnsymbol{footnote}}
\begin{titlepage}
\begin{flushright}
DFTT-49/2009\\
LTH 836
\end{flushright}
\vskip .7in
\begin{center}
{\Large \bf Current Exchanges for Reducible Higher Spin Multiplets
and Gauge Fixing } \vskip .7in {\large A. Fotopoulos
$^a$\footnote{e-mail: foto@to.infn.it} and M. Tsulaia
$^b$\footnote{e-mail: tsulaia@liv.ac.uk},}\vskip .2in {$^a$
Dipartimento di Fisica Teorica dell'Universit\`a di Torino and
INFN Sezione di Torino, via
P. Giuria 1, I-10125 Torino, Italy\\
\vskip .2in $^b$
 Department of Mathematical Sciences,
        University of Liverpool, Peach Street,
                Liverpool L69 7ZL, United Kingdom}

\begin{abstract}
 We compute the current exchanges between triplets of higher spin
 fields which describe reducible representations of the Poincare group.
 Through this computation we can extract the propagator of the
 reducible higher spin fields which compose the triplet. We show
 how to decompose the triplet fields into irreducible HS
 fields which obey Fronsdal equations, and how to compute
 the current-current interaction for the cubic couplings which
 appear in \cite{Fotopoulos:2007yq} using the decomposition into
 irreducible modes. We compare this result with the same
 computation using a gauge fixed (Feynman) version of the triplet
 Lagrangian which allows us to write very simple HS propagators
 for the triplet fields.

\end{abstract}

\end{center}

\vfill

\end{titlepage}

\tableofcontents

\section{Introduction}\label{Intro}

The theory of interacting massless and massive higher spin (HS)
fields (see \cite{Bekaert:2005vh}--\cite{Fotopoulos:2008ka} for
reviews) is attracting growing interest. Until now the consistent
interaction vertices for the massless and massive higher spin
fields, both on flat and constant curvature backgrounds, have been
obtained in frame - like \cite{Fradkin:1986qy} and in metric-like
\cite{Fronsdal:1978rb}-- \cite{Manvelyan:2009vy} formulations.
However, the studies in this directions are far from being
complete. The most challenging problems are to build the complete
systematics of the interacting higher spin fields and understand
the possible role  and connection of these kind of theories with
string and M- Theory. One particular problem involves further
study of the cubic interactions which have already been
constructed in various approaches in order to obtain effective
actions which contain  interaction terms of an order higher than
cubic (see e.g. \cite{Bengtsson:2006pw}).

In the present paper we shall follow  the covariant BRST
formulation of the interacting higher spin fields
\cite{Buchbinder:2006eq} (see also \cite{Bengtsson:1987jt} for the earlier
 work in this direction). As a  first step in this set-up one
constructs a BRST charge and the BRST-invariant free Lagrangian
(see \cite{Francia:2002aa} -- \cite{Bastianelli:2009vj} for other
gauge-invariant descriptions for massive and massless higher spin
fields)
 which  describes the propagation
of symmetric higher spin modes either on flat or $AdS$ space. As a
result of the gauge-invariant formulation, the free Lagrangian
contains a number of auxiliary fields. The total system of fields
is called a ``triplet'' for the case of  reducible massless symmetric
representations of Poincare or Anti de Sitter group
\cite{Ouvry:1986dv}--\cite{Sorokin:2008tf}, or ``generalized
triplet'' for the case of reducible representations of the
Poincare group with  mixed symmetry  \cite{Sagnotti:2003qa}.
 The second step is to make a consistent
nonlinear deformation of the quadratic Lagrangian and of the
abelian gauge transformations by building the BRST-invariant cubic
vertex. The extension of this method to the case of massive higher
spin fields is straightforward, the only difference being that one
has to use the BRST charge for massive reducible representations
of the Poincare group \cite{Hussain:1988uk}--\cite{Burdik:2000kj}.
An advantage of this approach is that in contrast to the BRST
charge describing the propagation of irreducible higher spin modes
\cite{Pashnev:1998ti}-- \cite{Buchbinder:2005ua}, the BRST charge
for triplets and generalized triplets
 has a  much simpler
form. This in turn simplifies the problem of finding  the
BRST-invariant cubic interaction vertex for either  massless or
massive fields.

However, it is not clear how far one can pursue the study of
interactions between reducible representations in order to build
complete systematics or at least to achieve a better understanding
of interacting higher spin theory. Therefore, a study of
interactions  between the fields which belong to the irreducible
representations of Poincare and $AdS$ groups are of  extreme
importance.
 There are several reasons for this: apart from
the fact that the original interacting higher spin theory
has been constructed for {\it irreducible} modes in the frame-like
approach \cite{Fradkin:1986qy}, a BRST charge describing a massive
triplet or a generalized triplet on an  $AdS$ space has
 not been constructed yet\footnote{One can construct, however,
the Lagrangian describing an interaction between a higher spin modes and massive
scalars from the one describing the massless fields
\cite{Fotopoulos:2007yq}}.
 Furthermore, a computation of four point scattering amplitudes, or Witten diagrams in the case
of an $AdS$ space for reducible higher spin modes can be rather
complicated because of the presence of nonphysical pure gauge
degrees of freedom in corresponding Lagrangians. Obviously these
degrees of freedom should be gauged away in order to build a
consistent perturbation theory. In other words, the pure gauge
degrees of freedom, which simplify the structure of the free
Lagrangian and the interaction vertexes, cause difficulties when
analyzing Feynman diagrams and scattering amplitudes and the main
goal of the present paper is  to address this problem.

 To summarise: our strategy is to start from
the Lagrangian describing interacting reducible representations of
the higher spin modes, since the construction of these kind of
gauge-invariant Lagrangians is much simpler than for irreducible
ones. As a second step  in order to build the perturbation theory
we extract the corresponding propagators for irreducible higher
spin modes from the Lagrangian describing massless reducible
higher spin fields. This program turns out to be technically
rather complicated and we describe it in great detail. As an
application of this procedure we consider the problem of
current-current exchange for the case where reducible higher spin
modes are coupled to scalar fields. We leave the application of
the technique developed in this paper for the case of more
complicated systems when one has  interactions between infinite
number of triplets, for  further study.

We would like to point out that our results described in Section
\ref{Decomp} for diagonaliziation of the  Lagrangian which
contains reducible fields in terms of Fronsdal Lagrangians for
irreducible fields, has been checked explicitly only up to spin
$4$. Nevertheless we have provided an ansatz which we believe that
it diagonalizes the triplet Lagrangian for arbitrary spin-s. Based
on the non-trivial nature of the diagonalization procedure for
spin $4$ as well as some qualitative features, which appear in the
comparison of these results with those of Section \ref{CET},  we
 suggest that the procedure proposed for the
decomposition of the Lagrangian describing ``triplets'' into the
Fronsdal Lagrangians for irreducible fields is correct.

\setcounter{equation}0\section{Higher Spin Triplets: Notation and
Conventions}\label{Not} In this Section we shall briefly summarise
the technique of building BRST-invariant cubic vertexes
\cite{Buchbinder:2006eq}.  We shall explain it on an example of
massive triplets which, despite being a very simple generalization
of the vertex given in  \cite{Fotopoulos:2007nm} has not been
presented elsewhere\footnote{A related discussion with respect to
the high energy limit of Open String Field Theory appears in
\cite{Bonelli:2003kh}}.

Let us start from the massless triplet in ${\cal D}+1$ space--time
dimensions. To this end we introduce an auxiliary Fock space
spanned by oscillator and ghost variables
\begin{equation}\label{B4}
[\alpha_M, \alpha_N^{+} ] =  \eta_{M N}, \quad \{ c^{+}, b \} = \{
c, b^{+} \} = \{ c_0 , b_0 \} = 1\,,
\end{equation}
and the vacuum in the Hilbert space is defined as:
\begin{equation}\label{B5}
\alpha_M |0\rangle =  0, \quad
 c|0\rangle =  0   , \qquad
b|0\rangle\ = \ 0, \quad b_0|0\rangle\ = \ 0.
\end{equation}
Obviously, one can consider an arbitrary number of these
oscillators, thus describing reducible representations of the
Poincare group with mixed symmetry \cite{Sagnotti:2003qa}.
Although the generalization to this case is straightforward we
shall consider  only totally symmetric representations. The
corresponding BRST charge has the form:
\begin{equation}\label{hebrst}
 Q = c_0 {\tilde l_0} + c^+ {\tilde l} + {c \tilde l^+} - c^+ c b_0
\end{equation}
with $\tilde l_0= p^M p_M$, $\tilde l = \alpha^M p_M$, $p_M = -i
\partial_M$.

The  functional (named ``triplet'' \cite{Francia:2002pt}) which
contains  both physical reducible representations of the Poincare
group with arbitrary integer spins and auxiliary nonphysical
fields is the most general expansion in terms of the ghost
variables with the ghost number zero
 \begin{equation}
\label{Phifield} |\Phi  \rangle = |\phi_1\rangle + c_0
|\phi_2\rangle=  |\varphi \rangle + c^+ \ b^+\ |d\rangle + c_0\
b^+\ |c\rangle \nonumber
\end{equation}
and the component fields are given by:
\begin{eqnarray}\label{Phiexp}
&& |\varphi\rangle = \frac{1}{s!}\, \varphi_{M_1 \ldots M_s}(x)
\alpha^{M_1+} \ldots \alpha^{M_s+}\;
|0\rangle \nonumber \nonumber \\
&&  |d \rangle = \frac{1}{(s-2)!}\, D_{M_1 \ldots M_{s-2}}(x)
\alpha^{M_1+}
\ldots \alpha^{ M_{s-2}+}  \; |0\rangle \ , \nonumber \\
&& |c \rangle \ = \ \frac{-i}{(s-1)!}\, C_{M_1 \ldots M_{s-1}}(x)
\alpha^{M_1+}  \ldots \alpha^{ M_{s-1}+}
 \; |0\rangle \ .
\end{eqnarray}

Furthermore, one can perform a dimensional reduction to ${\cal D}$
dimensions thus describing a massive theory in one dimension lower
\cite{Hussain:1988uk}--\cite{Burdik:2000kj}. The
corresponding BRST charge
\begin{equation}\label{hebrstm}
 Q = c_0 { l_0} + c^+ { l} + {c  l^+} + c_0 m^2 +
 c^+ m \alpha_D + c m \alpha_D^+ - c^+ c b_0, \quad l_0 = p^\mu p_\mu, \quad l= \alpha^\mu p_\mu,
\end{equation}
contains the constant mass parameter $m$ and therefore all fields
in the triplet have the same value of  mass.However one can make
 the mass parameter oscillator dependent
\cite{Pashnev:1997rm} thus considering a Regge trajectory similar
to the one present in the bosonic string theory. The construction
of the interaction vertex in this case is much more involved and
we shall not consider this interesting possibility here. Having
constructed the nilpotent BRST charge one can write the
BRST-invariant free Lagrangian
\begin{equation} \label{LM}
{\cal L}= \int dc_0  \langle \Phi|  Q|\Phi  \rangle
\end{equation}
which is invariant under the gauge transformations
\begin{equation} \label{GTM}
\delta | \Phi  \rangle = Q | \Lambda  \rangle, \quad |\Lambda
\rangle = b^+ |\lambda\rangle, \quad |\lambda \rangle \ = \
\frac{i}{(s-1)!}\, \lambda_{M_1 \ldots M_{s-1}}(x) \alpha^{M_1+}
\ldots \alpha^{ M_{s-1}+}
 \; |0\rangle \
\end{equation}
The free equations of motion and gauge transformation rules for
the massive triplet can be easily obtained from (\ref{Phiexp}),
(\ref{hebrstm}), (\ref{LM}) and (\ref{GTM}) after making the
decomposition $\alpha_M^+ \rightarrow (\alpha_\mu^+, \alpha_D^+)$
\begin{equation}\label{trf1m}
(l_0 +m^2)|\varphi \rangle = (l^+ + m \alpha_D^+)|c \rangle
\end{equation}
\begin{equation}\label{trf2m}
(l_0 +m^2) |d \rangle = (l + m \alpha_D)|c \rangle
\end{equation}
\begin{equation}\label{trf3m}
|c \rangle = (l^+ + m \alpha_D^+) |d \rangle - (l +m
\alpha_D)|\varphi \rangle
\end{equation}
while the gauge transformation rule (\ref{GTM}) gives
\begin{equation}\label{GTt1m}
\delta |\varphi \rangle = (l^+ +m \alpha_D^+)|\lambda \rangle,
\quad \delta |d \rangle = (l  + m \alpha_D)|\lambda \rangle, \quad
\delta |c \rangle = (l_{0} +m^2)|\lambda \rangle.
\end{equation}

In order to describe  cubic interactions  one introduces
 three copies ($i=1,2,3$) of the Hilbert space
defined above, as in bosonic Open String Field Theory
\cite{Gross:1986ia}. Then the Lagrangian has the form
\begin{equation} \label {LIBRST}
{\cal L} \ = \ \sum_{i=1}^3 \int d c_0^i \langle \Phi_i |\, Q_i \,
|\Phi_i \rangle \ + g( \int dc_0^1 dc_0^2  dc_0^3 \langle \Phi_1|
\langle \Phi_2|\langle \Phi_3||V \rangle + h.c)\,, \end{equation}
where $|V\rangle$ is the cubic vertex and $g$ is a
 coupling
constant. The Lagrangian (\ref{LIBRST}) is  invariant up to the
first order in the coupling constant $g$ with respect to the
nonabelian gauge transformations
\begin{equation}\label{BRSTIGT1}
\delta | \Phi_i \rangle  =  Q_i | \Lambda_i \rangle  - g \int
dc_0^{i+1} dc_0^{i+2}[(  \langle \Phi_{i+1}|\langle \Lambda_{i+2}|
+\langle \Phi_{i+2}|\langle \Lambda_{i+1}|) |V \rangle] \,,
\end{equation}
provided that the vertex $|V\rangle$ satisfies the BRST invariance
condition
\begin{equation}\label{VBRST}
\sum_i Q_i |V \rangle=0\,.
\end{equation}
Further on, in order to simplify equations in the rest of this
section we introduce bilinear combinations of the oscillators
\begin{equation}\label{Defab}
\gamma^{+,ij}=c^{+,i} b^{+,j}, \quad \beta^{+,ij}=c^{+,i}
b^j_{0}\, \quad M^{+, ij}= \frac{1}{2} \alpha^{+,\mu, i}
\alpha^{+,\mu,j}
\end{equation}
which have ghost number zero. Further we take an ansats for the
vertex
\begin{equation}\label{KOmod}
|V \rangle=  V^{1} \times V^{2}|-\rangle_{123},
\end{equation}
$$
 \quad |-\rangle_{123}= c_0^1 c_0^2 c_0^3 |0 \rangle_1 \otimes |0 \rangle_2 \otimes |0 \rangle_3
$$
with
\begin{equation}\label{KOansatz}
V^{1}= exp\ (\ Y_{ij} l^{+,ij} + Z_{ij} \beta^{+,ij} + U_{ij} m^i
\alpha_D^{+,j}\ )\,,
\end{equation}
$$
 V^{2}=exp\ (\ S_{ij} \gamma^{+,ij} +
P_{ij} M^{+, ij} + R_{ij}\alpha_D^{+,i} \alpha_D^{+,j} \ ),
$$
where $P_{ij}=P_{ji}$, $R_{ij}=R_{ji} $. We have also assumed that
$m_1=m_2=m_3$. However, this requirement is not a necessity  and one
can still find a solution when this requirement is relaxed.
 Putting this ansatz into the BRST invariance condition
and using momentum conservation $p_\mu^1+ p_\mu^2 + p_\mu^3=0$ one
can obtain a solution for $Y^{rs}, U^{rs}$ and $Z^{rs}$
\begin{equation}\label{KOsolution}
Z_{i,i+1}+Z_{i,i+2}=0
\end{equation}
$$
Y_{i,i+1}= Y_{ii}-Z_{ii} -1/2(Z_{i,i+1}-Z_{i,i+2})
$$
$$
Y_{i,i+2}= Y_{ii}-Z_{ii} +1/2(Z_{i,i+1}-Z_{i,i+2}).
$$
$$
Z_{i,i}+ Z_{i,i+1}+ Z_{i, i+2}= U_{i,i}+U_{i+1,i}+U_{i+2,i}
$$
\begin{eqnarray}\label{KOmsol}
&& S_{ij}= P_{ij}=R_{ij}=0 \qquad i\neq j \\
&& R_{ij}- S_{ii}=0 \qquad i=1,2,3 \\
&& P_{ii} - S_{ii}=0 \qquad i=1,2,3 \nonumber
\end{eqnarray}
In what follows we will assume cyclic symmetry in the three Fock
spaces which implies along with (\ref{KOsolution}) and
\begin{equation}\label{cyclic}
Z_{12}=Z_{23}=Z_{31}=Z_a, \quad Z_{21}=Z_{13}=Z_{32}=Z_b=-Z_a
\end{equation}
$$
U_{12}=U_{23}=U_{31}=U_a, \quad U_{21}=U_{13}=U_{32}=U_b
$$
$$ Y_{12}=Y_{23}=Y_{31}=Y_a, \quad
Y_{21}=Y_{13}=Y_{32}=Y_b
$$
$$
Y_{ii}=Y, \quad \ Z_{ii}= Z,  \quad P_{ii}=P, \quad
S_{ii}=S, \quad R_{ii}= R, \quad
$$
$$
S=P=R
$$
Choosing the value of the parameter $S$ to be equal to $1$ one can
make the above solution exact to all orders in the coupling
constant in complete analogy with \cite{Fotopoulos:2007nm}. It can
be checked directly that this solution belongs to nontrivial
cohomologies of the BRST charge (\ref{hebrstm}) and thus can not
be obtained via the field redefinitions from the free
Lagrangian\footnote{ The nontrivial cohomology means that the
vertex cannot be written in the form $|V\rangle = \tilde{Q} | W
\rangle $, where $| W \rangle$ an arbitrary functional having
ghost number $-2$ (see \cite{Buchbinder:2006eq} for details). Such interactions are generated by field
redefinitions of the form $| \varphi_i\rangle = \langle \varphi_{i+1},
\varphi_{i+2} |  |W\rangle $. Actually if we drop the
requirement for cyclic symmetry the most general field
redefinition would generate an interaction of the form $ |\delta
V\rangle = Q_1 |W_1\rangle + Q_2 |W_2\rangle + Q_3 |W_3\rangle $,
which corresponds to the case when we redefine the three fields
separately in the free Lagrangian using three independent
functionals $|W_i\rangle$.
Therefore in practice when the cyclic symmetry of the vertex is not required in order to exclude all vertexes
 which can be obtained from the free Lagrangian via the field redefinitions one needs to check that the solution of cohomologies
 of the BRST charge $\tilde Q$ does not have the form $Q_1 |W_1 \rangle + Q_2 |W_2 \rangle + Q_3 |W_3 \rangle$.}
 Let us also note that one can make the
triplet matrix valued, and consider a theory with a nonabelian
gauge group in complete analogy with the string theory.

As an alternative example one can consider a different condition
on the mass parameters; in particular $m_1+m_2+m_3=0$. This case
will correspond to a dimensional reduction of the vertex given in
\cite{Fotopoulos:2007nm}. In this case  parameters $U_{ij}$ will
obey exactly the same conditions as the parameters $Y_{ij}$.

An important point is that one can consistently put the mass
parameter(s) $m_i$ equal to zero and decouple oscillators
$\alpha_D^+$ which correspond to the compact dimension. In this
way one recovers the interacting system of massless triplets described
in  \cite{Fotopoulos:2007nm}. As for the case of interacting
massless triplets, the case of interacting massive triplets
requires an infinite number of them in order to ensure the
exactness of the vertex in all orders in coupling constant $g$.

The next natural step is to construct the full perturbation theory
for this model of interacting higher spins. To this end one needs
to gauge fix the action in order to avoid  propagation of the pure
gauge degrees of freedom and extract the propagators of individual
physical higher spin modes. This program turns out to be
technically involved and as a first step instead of building the
perturbation theory for this model, we shall consider a system of
interacting triplets where their number is finite. As we mentioned
above in this case the vertex is no longer exact to all orders in
the coupling constant and therefore there is a certain freedom in
the definition of the coupling constants. These coupling constants
can be presumably fixed in the full interacting theory which in
principle can be different from the solution described above,
since one can not claim that this solution is unique.

Therefore in the following, let us consider the simple case of the
system of one massless triplet which describes spins $s,
s-2,...,1/0$ interacting with two scalar fields
\cite{Fotopoulos:2007yq}. The general solution for a gauge
invariant Lagrangian to the lowest order in $g$ can  in principle
be deduced from (\ref{KOsolution}) if one drops the requirement
for cyclic symmetry and sets $m_i=0$. The exactness of the vertex
to all orders in $g$ is also no longer required, so one can
consider a finite number of interacting triplets.
 If we keep the
masses of the two scalars nonzero the interaction vertex remains
the same as for the massless case, but the free Lagrangian is that of
massive scalar triplets. This is the equivalent model to
the one considered in \cite{Bekaert:2009ud}, where the scalars
play the role of matter charged under HS gauge fields. These higher spin fields  gauge
the rigid symmetries of the  free action for the scalars. One can also make
a deformation of this solution for an $AdS$ space but we shall not
consider this possibility here.
 In components the free part of the Lagrangian of (\ref{LIBRST}),
for a spin-s triplet can be written
\begin{eqnarray}\label{Lcomp}
{\cal L} &=& - \, \frac{1}{2}\ (\partial_\mu \varphi)^2 \ + \ s\,
\partial \cdot \varphi \, C \ + \ s(s-1)\, \partial \cdot C \, D \ \nonumber \\
&+&  \ \frac{s(s-1)}{2} \ (\partial_\mu D)^2 \ - \ \frac{s}{2} \,
C^{\; 2} \ ,
\end{eqnarray}
where we have rescaled all fields of the triplet by a factor
$\varphi_s \to \sqrt{s!}\  \varphi_s$ as can be easily seen by
computing the bracket $\langle \Phi | Q | \Phi \rangle$ using the
definitions of (\ref{Phiexp}). The free equations of motion for a
triplet on a flat background are
\begin{eqnarray}\label{eqm1}
&& \Box \; \varphi \ = \ \partial C  \nonumber \ , \\
&& C = \partial \cdot \varphi - \partial D \nonumber \ , \\
&& \Box \; D \ = \ \partial \cdot C \  \ , \label{triplet}
\end{eqnarray}
along with the gauge transformations
\begin{equation}\label{deltaphi}
\delta \phi = \partial \lambda, \quad \delta C = \Box \lambda,
\quad \delta D = \partial \cdot \lambda.
\end{equation}
Here, as usual, $\partial \cdot$ denotes the divergence and $\partial$ denotes the symmetrized
derivative without contraction of indexes.
Using the expression of the vertex in \cite{Fotopoulos:2007yq}, or
equivalently setting $m_i=0$ and dropping the cyclicity
requirement in (\ref{KOmod})-(\ref{cyclic}), we can write the
cubic interaction for two scalars and one arbitrary HS triplet
\begin{equation}\label{ILW}
{\cal L}_{int}^{00s}=   \sum^{[{s \over 2}]}_{q=0} \
\frac{N_{s-2q}}{(2q)!! (s-2q)!} \  {\cal W}_s^{q} \cdot
J^{1;2}_{s-2q}\ + \ h.c.\,,
\end{equation}
where ${\cal W}_s^q$ is defined in \cite{Francia:2002pt}
\begin{equation}\label{hatS}
{\cal W}_s^{q}= \varphi_s^{[q]}-2q \ D_{s-2}^{[q-1]}\,, \,\,\,\delta {\cal
W}^{q}_s=
\partial \Lambda^{[q]}_{s-1} \,,
\end{equation}
and $\varphi^{[q]}_s$ is the qth trace of the tensors $\varphi_s$
of rank-s. The currents are defined as \cite{Berends:1985xx}
\begin{equation}\label{Jp}
J^{1;2}_{s-2q}=\sum_{r=0}^{s-2q} \ \bn{s-2q}{r} \ (-1)^r \
(\partial^{\mu_1} \dots
\partial^{\mu_r}
\phi_1)\ (\partial^{\mu_{r+1}} \dots
\partial^{\mu_{s-2q}}\phi_2)\,
\end{equation}
and $N_{s-2q}$ is an undetermined constant which probably gets
fixed once we have the fully consistent,
interacting HS theory
to all orders in the coupling constant.
 That is, we expect that
closing the algebra of gauge transformations  and gauge invariance
of higher order interactions will constrain these coefficients.
Their precise values are not important for our present discussion
and we will not use them any more, but we will assume that they
are non-vanishing and therefore consistent interactions of the
type in (\ref{ILW}) do exist in the fully-gauge invariant theory.
Moreover the explicit form of the currents is not needed in what
follows. Since we shall consider current-current interactions for
external currents with intermediate HS states propagating, the
only property we shall use for our computations is their
conservation. Nevertheless, if one  considers a scattering process
between dynamical scalar fields, the explicit form of the currents
is needed \cite{Bekaert:2009ud}.

Our goal in the following sections is to compute the
current-current interaction between the currents (\ref{Jp}) using
two methods. First, by decomposing ${\cal W}_s^q$ into irreducible
fields; and  second  by directly using the ${\cal W}_s^q$
propagator in a particular gauge. This way we will deduce the
propagator of the triplet fields in (\ref{Lcomp}) which
to our knowledge has not
been considered elsewhere.

\setcounter{equation}0\section{~Decomposition ~of ~Higher ~Spin ~Triplets ~Into
~Irreducible ~Higher ~Spin ~Fields ~and ~Current-Current
~Interactions}\label{Decomp}

In order to compute scattering amplitudes using Feynman rules we will need
to decompose the triplets into irreducible modes and apply this
decomposition to the free Lagrangian (\ref{Lcomp}). We expect that the
Lagrangian for the triplet
 will become a sum of Fronsdal Lagrangians for
irreducible higher spin states of spin $s, s-2,...,1/0$.   This
decomposition is a long and difficult task which to our knowledge
has not been presented elsewhere. We will take an interacting Lagrangian to
be of the form (\ref{ILW}) with currents given by linear
combinations of those in (\ref{Jp}). Once we have completed this
task we will compute the current exchanges using the same methods
as in \cite{Francia:2007qt}.

\subsection{Decomposition Of ${\cal W}$ States In Terms Of
Irreducible Modes}\label{irr}

In this subsection we will demonstrate how we can decompose the
fields ${\cal W}^q_s$ in terms of individual (Fronsdal) higher
spin modes $\Psi$. We use the ${\cal W}^q_s$ fields since their
gauge transformation has the simple form (\ref{hatS}).
Nevertheless, ${\cal W}$ transform with a tracefull gauge
parameter and off shell their double trace is not zero, unlike the
double trace of the irreducible higher spin mode which appears in
 the Fronsdal  description
\cite{Fronsdal:1978rb}.
Moreover, the equations of motion for ${\cal W}_s^q$ are not
decoupled among each other
\begin{equation}\label{eqmW}
{\cal F}{\cal W}^q_s=\Box {\cal W}^q_s-\partial \partial \cdot
{\cal W}^q_s + \partial^2 {({\cal W}^{q}_s)}^\prime=
\partial^2 {\cal W}^{q+1}_s
\end{equation}
where ${\cal F}$ is the Fronsdal operator. After complete gauge
fixing the triplet describes irreducible HS fields with spins $s,
s-2 \dots 1/0$.  These physical modes correspond to the on-shell
modes of ${\cal W}^q_s$ \cite{Francia:2002pt,Fotopoulos:2008ka}.
Notice that setting in (\ref{eqmW}) ${\cal W}^q_s=0, q \geq 1$ we
recover the Fronsdal equations of motion for an irreducible HS
field of spin-s
 and all lowest spin fields effectively dissapear.
Our goal is to extract irreducible  Fronsdal fields from the
triplet Lagrangian,  which are double traceless (off shell) and
transform with a traceless gauge parameter.

Let us demonstrate with a few low spin examples our decomposition
method and then we will give the general formula. It is more
convenient to work with the Lagrangian after we have eliminated
the auxiliary field $C_{s-1}$
\begin{eqnarray}\label{Lcomp2}
{\cal L} &=& - \, \frac{1}{2}\ (\partial_\mu \varphi)^2 \ + \
\frac{s}{2} \,
(\partial \cdot \varphi)^2 \ + \ s(s-1)\, \partial \cdot \partial \cdot \varphi \, D \ \nonumber \\
&+& \ s(s-1) \ (\partial_\mu D)^2 \ + \ \frac{s(s-1)(s-2)}{2} \,
(\partial \cdot D)^2  \ .
\end{eqnarray}
Let us note, that the non dynamical field $C$ naturally appears as a result  of a general expansion
of the functional $| \Phi \rangle$ in terms of the ghost variables. Moreover its presence is required by the form of the gauge transformations (\ref{GTM})
since the BRST charge (\ref{hebrst}) when acting on the parameter of gauge transformations $| \Lambda \rangle$  gives rise to a term $-c_0 b^+ \Box | \lambda \rangle$
proportional to the combination $c_0 b^+$.  In principle one can consider the system with the constrained parameter of gauge transformations
$\Box \lambda =0$.
Otherwise one can
  notice that the gauge transformation rule for the field $C$ coincides with the one for
$\partial \cdot \varphi - \partial D$. Therefore one can express the field $C$ in terms
of the fields $\varphi$ and $D$ before constructing the BRST invariant Lagrangian and thus consider
a ``doublet'' formulation for the free reducible higher spin modes.
However in a case of interacting triplets, depending on a particular vertex under consideration
it is not so easy in general to guess a correct form of the expression of  fields $C_i$ in terms of
the fields $\varphi_i$ and $D_i$. Although in the particular example of a  cubic interaction
considered in the present paper one can still  do so (the gauge transformation rule for  $\phi$, $C$ and $D$ fields
which describe the higher spin triplet does not change, whereas the scalars do not bring about $C$ fields \cite{Fotopoulos:2007yq})  we keep the field $C$
from the beginning in order to keep a systematic BRST approach for interacting triplets.

The Lagrangian above can be written in a fully symmetrized form
\begin{eqnarray}\label{Lsym}
{\cal L} &=& - \, \frac{1}{2(s+1)}\ (\partial \varphi)^2 \ + \ s\,
(\partial \cdot \varphi)^2 \ - \ s\, (\partial \cdot \varphi) \, (\partial D) \ \nonumber \\
&+& \ s \ (\partial D)^2 \ - \ \frac{s(s-1)(s-2)}{2} \, (\partial
\cdot D)^2  \,
\end{eqnarray}
where we have used the following identity for any symmetric field
of spin s
\begin{equation}\label{fDiden}
(\partial_\mu \varphi)^2= {1\over s+1} (\partial \varphi)^2 - s
(\partial \cdot \varphi)^2.
\end{equation}
We will also use  the following identities and conventions
described  in \cite{Francia:2002pt}
\begin{eqnarray}\label{etak}
&& \left( \partial^{\; p} \ \varphi  \right)^{\; \prime} \ = \
\Box \
\partial^{\; p-2} \
\varphi \ + \ 2 \, \partial^{\; p-1} \  \partial \cdot \varphi \ +
\
\partial^{\; p} \
\varphi {\;'} \ , \nonumber \\
&& \partial^{\; p} \ \partial^{\; q} \ = \ \left( {p+q} \atop p
\right) \
\partial^{\; p+q} \ ,
\nonumber \\
&& \partial \cdot  \left( \partial^{\; p} \ \varphi \right) \ = \
\Box \
\partial^{\; p-1} \ \varphi \ + \
\partial^{\; p} \ \partial \cdot \varphi \ ,  \\
&& \partial \cdot  \eta^{\;k} \ = \ \partial \, \eta^{\;k-1} \ , \nonumber \\
&& \left( \eta^k \, T_{(s)} \,  \right)^\prime \ = \, \left[ \, d
\, + \, 2(s+k-1) \,  \right]\, \eta^{k-1} \, T_{(s)} \ + \ \eta^k
\,  T_{(s)}^\prime \ \nonumber  ,
\end{eqnarray}
The symmetrization notation is the one of \cite{Francia:2002pt}.
Finally, for completeness we give the Fronsdal Lagrangian
\cite{Fronsdal:1978rb} for an irreducible higher spin field $\Psi$
\begin{eqnarray}\label{Fronsdal}
{\cal L}&=& -{1\over 2} (\partial_\mu \Psi)^2 + \frac{s(s-1)}{4} {(\partial_\mu \Psi^\prime)}^2
+ \frac{s}{2}{(\partial \cdot \Psi)}^2 \\ \nonumber
&& + \frac{s(s-1)}{2} \Psi^\prime (\partial \cdot \partial \cdot \Psi)
+\frac{s(s-1)(s-2)}{8}(\partial \cdot \Psi^\prime) (\partial \cdot \Psi)
\end{eqnarray}

\paragraph{Spin-2}
We make the ansatz
\begin{eqnarray}\label{s2}
{\cal W}_2^0 &=&\varphi_{\mu \nu}= \Psi_{\mu \nu}- A \eta_{\mu \nu} \Psi \nonumber \\
{\cal W}_2^1 &=& \varphi^{'}-2D=\Psi
\end{eqnarray}
By direct substitution in (\ref{Lcomp2}) we can easily see that
the $\Psi_{\mu \nu}, \Psi$ fields decouple
 for $A= -{1 \over d-2}$ and the Lagrangian becomes
 \begin{equation}\label{Ls2}
 {\cal L}= -{1 \over 2} (\partial_\mu \Psi_{\rho \sigma})^2 +
 (\partial_\nu \Psi^\nu_\mu)^2 + \Psi^{'}
 \partial_{\mu}\partial_{\nu} \Psi^{\mu \nu} + {1\over 2}
 (\partial_\mu \Psi^{'})^2 - {1 \over 2(d-2)} (\partial_\mu \Psi)^2.
 \end{equation}
 The Lagrangian above describes a massless spin-2 particle and a
 scalar. We note that the normalization of the kinetic term of the
 scalar is not canonical and this will have to be taken into
 account when computing the current exchanges since the residue of the propagator will not be the standard one.
 Now we can solve
 (\ref{s2}) for $\Psi_{\mu \nu}$ and $\Psi$ in terms of the fields
  ${\cal W}^0_{2; \mu \nu}$ and ${\cal W}^1_2$ the solutions is

\begin{eqnarray}\label{Ws2}
\Psi_{\mu \nu} &=&{\cal W}_{2;\mu \nu}^0  -{1\over d-2} \eta_{\mu\nu} {\cal W}_2^1 \nonumber\\
\Psi &=& {\cal W}_2^1.
\end{eqnarray}

\paragraph{Spin-4}
The above spin-2 result suggests that we expand the $\Psi_n$
fields in terms of ${\cal W}^q_s$. This is a crucial observation
based on the fact that: i) ${\cal W}_s^q$ transform like
irreducible modes but with traceful  parameters of gauge
transformations ii) the fields $\Psi_n$, since they are
irreducible, should be double traceless $\Psi_n^{''}=0, \ n>3$,
and should transform with traceless gauge parameter. We make the
ansatz
\begin{eqnarray}\label{s4}
\Psi_4 &=& {\cal W}_4^0 + A \eta {\cal W}_4^1 + B \eta^2 {\cal W}_4^2 \nonumber \\
\Psi_2 &=& {\cal W}_4^1 + C \eta {\cal W}_4^2 \\
\Psi_0 &=& {\cal W}_4^2 \nonumber
\end{eqnarray}
where $\Psi_n$ denotes irreducible higher spin modes with spin
$n$. From (\ref{s4}) and (\ref{hatS}) one can see that, the gauge
transformation rule for the field  $\Psi_4$ is $\delta
\Psi_4=\partial {\tilde \Lambda}_3$ with $\tilde {\Lambda}_3$
given by
\begin{equation}
{\tilde \Lambda}_3= \Lambda_3+A\eta \Lambda'_3.
\end{equation}
If we demand that $\tilde {\Lambda}_3$  is a traceless tensor of
the third rank, as required by  \cite{Fronsdal:1978rb} we get
$A=-{1\over d+2}$. However the requirement of the ``proper'' gauge
transformation does not fix all coefficients in (\ref{s4}). In
order to fix the remaining coefficients one uses the condition
$\Psi_4''=0$  and
\begin{equation}\label{Wtr}
({\cal W}^q)'= {1 \over q+1} \varphi^{[q+1]} + {q\over q+1} {\cal
W}^{q+1}.
\end{equation}
to obtain
\begin{equation}\label{s4con}
A=-{1\over d+2} \qquad B= {1\over d(d+2)}.
\end{equation}
So the double tracelessness condition allows us to fix most of the
coefficients in (\ref{s4}) and moreover guarantees that $\Psi_4$
transforms as an irreducible field of spin 4  in agreement with \cite{Fronsdal:1978rb}.
We note
that we could have tried to put a ${\cal W}^{0'}$ term in the
expansion of $\Psi_2$ but this would give gauge transformation
terms of the form $\partial \cdot \Lambda_3$ which are not
appropriate for an irreducible mode. The remaining coefficients in
(\ref{s4}) cannot be fixed by the double tracelessness condition.
We assume that the decomposition of $\Psi_2, \Psi_0$ in terms of
${\cal W}^1_4$ and ${\cal W}^2_4$ is the same as the decomposition
of the spin-2 triplet in (\ref{Ws2}), that is we set $C=-{1\over
d+2}$.  This assumption will be verified later on when we will
demonstrate that this choice leads to a complete decoupling of the
$\Psi_n$ fields in the free Lagrangian. Of course we can solve
 the equations above for ${\cal W}$ or for $\varphi, D$ in
terms of for $\Psi_4, \Psi_2 $ and $\Psi$. The resulting
expressions are (the subscript in $\Psi_n$  denotes the rank of
the symmetric tensor field):
\begin{eqnarray}\label{Ws4}
{\cal W}_4^0 &=&  \Psi_4 +{1\over d+2} \eta \Psi_2 + {1\over d(d-2)} \eta^2 \Psi_0 \nonumber \\
{\cal W}_4^1 &=& \Psi_2 + {1\over d-2} \eta \Psi_0\\
{\cal W}_4^2 &=& \Psi_0 \nonumber
\end{eqnarray}
and
\begin{eqnarray}\label{phiDs4}
\varphi_4 &=& \Psi_4 + {1\over d+2} \eta \Psi_2 + {1\over d(d-2)} \eta^2 \Psi_0 \nonumber \\
D&=& {1\over 2} [ \Psi'_4 +{2\over d+2} \Psi_2 + {1\over d+2} \eta
\Psi'_2 +{2\over d(d-2)} \eta \Psi_0].
\end{eqnarray}

At this point we have managed to construct fields $\Psi_n$ which
satisfy the Fronsdal off-shell condition $\Psi_n^{''}=0$ but this
is not enough. We should check that the decomposition in
(\ref{s4}), or equivalently (\ref{phiDs4}), decompose the
Lagrangian (\ref{Lsym}) into a series of Fronsdal Lagrangians for
the irreducible fields $\Psi_n$. We demonstrate this explicitly
for the s=4 case in Appendix \ref{AppA}, where we show that all
cross-terms between $\Psi_n, \ n=0,2,4$ vanish.

\paragraph{spin-s}
The above arguments draw a clear strategy for finding the
decomposition for the general spin-s.\footnote{It should be clear
that the basis we have chosen for the decomposition in (\ref{ss})
is not the largest possible we could have constructed. In
principle the independent ``basis-vectors'' we could have written
are $\varphi, D$ and all their traces, a total of $s+1$ terms. We
have used instead only the $[{s\over 2}]+1$  linear combinations
of them given by ${\cal W}^q_s$. This is motivated by our
observation that the gauge transformations of ${\cal W}^q_s$ have
the proper form for giving irreducible fields gauge transformation
with traceless gauge parameters. It is plausible that our basis
might not be unique.}We assume an expansion of the form
\begin{equation}\label{ss}
\Psi_s= \sum_{q=0}^{[{s \over 2}]} \rho_q(d-2,s) \eta^q {\cal
W}_s^q.
\end{equation}
Imposing the double tracelessness condition it turns out that we
get more equations than parameters since the traces of ${\cal W}$
are not written in terms of ${\cal W}$ only (see \ref{Wtr}).

So the system seems over-constrained since we have to demand that
both the coefficients of $\varphi^{[q]}$ terms and ${\cal W}^q$
vanish after taking the double trace. Nevertheless, surprisingly, we find
a solution
\begin{equation}\label{rho}
\rho_q(d-2,s)= - {\rho_{q-1}(d-2,s) \over (d+2(s-q-2))}= {(-1)^q
(d+2(s-q-3))!! \over (d+2(s-3))!!}
\end{equation}
which appeared in \cite{Francia:2007qt}. In the same manner we can
show that in general
\begin{equation}\label{ss2}
\Psi_{s-2k}= \sum_{q=0}^{[{s \over 2}]-k} \rho_q(d-2,s-2k) \eta^q
{\cal W}_s^{q+k}
\end{equation}
are doubly traceless fields. Taking the gauge transformation of
this equation we can show, using $\rho_k(d-2,s)= \rho_k(d,s-1)$,
that $\delta \Psi_{s-2k}=\partial {\tilde \Lambda}_{s-2k}$ with
the traceless gauge parameter
\begin{equation}\label{tLambda}
{\tilde \Lambda}_{s-1-2k}= \sum_{q=0}^{[{s\over 2}]}
\rho_q(d,s-2k-1) \eta^q \Lambda^{[q+k]}_{s-1}
\end{equation}
 If one tries to invert these equations one gets a system of $[{s\over
2}]$ linear equations with a lower diagonal matrix. We write an
expansion of the form
\begin{equation}\label{Ws}
{\cal W}_s^{q}= \sum_{k=0}^{[{s \over 2}]-q} {\tilde
\rho}_k(d,s-2q) \eta^k \Psi_{s-2q-2k}.
\end{equation}
Generalizing equations (\ref{Ws2}) and (\ref{Ws4}) for the cases
$s=2,4,6$ we make an ansatz
\begin{equation}\label{trho}
\tilde{\rho}_k(d,s)= {(d+2(s-2k-2))!! \over (d+2(s-k-2))!!}.
\end{equation}
To verify our ansatz we insert (\ref{Ws}) in (\ref{ss}) and vice
versa. Then i.e. inserting the expansion of ${\cal W}^q_s$ into
the expansion of $\Psi_s$ we should get a Kroenecker delta on the
RHS of (\ref{ss}) as it is required by consistency with the LHS.
The same for the other way around. We finally get the condition
\begin{equation}\label{delta}
\delta_{0,u}= \sum_{n=0}^u \bn{u}{n} {\tilde \rho}_n(d,h)
\rho_{u-n} (d-2, h-2n)=\sum_{n=0}^u \bn{u}{n} {\tilde
\rho}_{u-n}n(d,h-2n) \rho_{n} (d-2, h).
\end{equation}
A direct computation with Mathematica gives a non-vanishing result
only for $u=0$, which implies the validity of the ansatz
(\ref{trho}) as a solution of (\ref{Ws}) for arbitrary spin $s$.
We have also checked it by hand up to spin $6$.
 Further on we need the expansion of $\varphi, D$ in terms
of $\Psi$ in order to verify that the irreducible modes decouple
among each other and to determine the normalization of the kinetic
term for each HS irreducible mode. These are given by
\begin{eqnarray}\label{fD}
\varphi &=& {\cal W}^0= \sum_{k=0}^{[{s \over 2}]}
\tilde{\rho}_k(d,s) \eta^k \Psi_{s-2k} \nonumber \\
D= {1\over 2} (({\cal W}^0)'- {\cal W}^1)&=& {1\over
2}\sum_{k=0}^{[{s \over 2}]-1} \tilde{\rho}_k(d,s) \eta^k
\Psi^{'}_{s-2k} + \sum_{k=1}^{[{s \over 2}]} \tilde{\rho}_k(d,s)
\eta^{k-1} \Psi_{s-2k}.
\end{eqnarray}
Now we can insert these expressions in the action (\ref{Lsym}) and
verify that all cross-terms vanish and therefore the fields
decouple. We have done this only up to spin 4 (see Appendix
\ref{AppA}) but we are confident that it works for all spins. In
any case taking as a fact the decoupling, the next thing to do is
to compute the normalization of the Fronsdal Lagrangian for each
irreducible higher spin field as it appears in the original
Lagrangian (\ref{Lcomp2}) after we insert the decomposition
(\ref{fD}). For this computation it is more convenient to use
(\ref{Lsym}). The key point is that we should look for the
normalization of the $(\partial_\mu \Psi_n)^2$ term for each
irreducible field. The standard normalization in Fronsdal
Lagrangian (\ref{Fronsdal}) is $-{1\over 2}$. In this manner we
get from (\ref{norm2}) the normalization for the propagator (the
inverse of the prefactor of  $(\partial_\mu \Psi_{s-2k})^2$ terms
multiplied by 2)
\begin{equation}\label{Q}
Q(s,k,d)= { 2^k k! (s-2k)!\over s!{\tilde \rho}_k(d,s)}.
\end{equation}

\subsection{The Current-Current Interaction }\label{CI}

In this subsection we will rewrite the interaction Lagrangian
(\ref{ILW}) in terms of the irreducible HS fields $\Psi_n$. We
will be interested in an
interaction term which contains a
 single ${\cal W}_s^q$.
The resulting Lagrangian term after we insert (\ref{Ws}) is
\begin{equation}\label{LWint}
{\cal L}(s,h) = {\cal W}_s^q \cdot J_{s-2q} = \sum_{k=0}^{[{s\over
2}]-q} \tilde{\rho}_k(d,s-2q) (\eta^k \Psi_{s-2q-2k}) \cdot
J_{s-2q}
\end{equation}
where $h=s-2q$. Using the identity (\ref{combi}) and
\begin{equation}\label{combiJ}
\eta^k\cdot J_{s-2q}= (2k-1)!! J^{[k]}_{s-2q}
\end{equation}
we get
\begin{equation}
{\cal L}(s,h) =   \sum_{k=0}^{[{s\over 2}]-q}
\tilde{\rho}_k(d,s-2q) {h! \over 2^k k! (h-2k)!} \Psi_{h-2k} \cdot
J^{[k]}_{h}.
\end{equation}
 The Lagrangian can be written as
\begin{equation}\label{ILP}
{\cal L}(s,h)= \sum_{k=0}^{h} \tilde{J}_{h-2k} \cdot \Psi_{h-2k}
\end{equation}
with
\begin{equation}\label{tJ}
\tilde{J}_{h-2k}=   \tilde{\rho}_{k}(d,h) {h! \over 2^k k! (h-2k)!
} J^{[k]}_{h}.
\end{equation}
The currents $\tilde{J}_{h-2k}$ are conserved since $J_{h}$ are
conserved for on-shell scalar fields in (\ref{Jp}). As explained
in \cite{Francia:2007qt} we can compute the current-current
interaction between those currents using a projector of the form\footnote{ The operator ${\cal P}$ is
polynomial of powers of
$\eta_{\mu \nu}$ and $\Pi_{\mu \nu}= \eta _{\mu \nu}- p_\mu
\bar{p}_\nu - p_\nu \bar{p}_\mu$ where $p^2=\bar{p}^2=0, \ p\cdot
\bar{p}=1$, which guarantees that only physical degrees of freedom
are exchanged between external currents. When acting on conserved
currents the expression of ${\cal P}$ simplifies and becomes
completely independent of $\bar{p}_\mu$, see (\ref{curexch}).}
${\cal P}^{h-2k}$, although the currents are not doubly traceless
as one would naively expect. It can be shown that despite this
apparent paradox, the correct number of physical modes is
exchanged when we compute the current-current interaction for
conserved currents just as in our case. Actually we can see that,
if we try to construct double traceless currents from the currents
of (\ref{Jp}) exactly as in (\ref{ss}), we get a current which is
not traceless conserved (traceless conserved current means that
its double trace and the traceless part of the divergence vanish
separately)
 as it is required by gauge invariance of the
total, free plus interaction, Lagrangian \footnote{Indeed gauge
variation of the free plus interacting Lagrangian leads to the
condition
\begin{equation} \delta {\cal L}= \int \Lambda \
\partial \cdot J = 0 \end{equation} which requires, for a double
traceless current J of spin s, that
\begin{equation}
\partial \cdot J= {1 \over d+2(s-3)} \ \eta (\partial \cdot J')
\end{equation}
since the gauge parameter is traceless for the gauge variation of
an irreducible field.}.

The expression for the current exchange of two conserved currents
coupled to the irreducible higher spin fields $\Psi_{h-2k}$ is
given by
\begin{equation}\label{curexch}
\tilde{J}_{h-2k} \cdot {\cal P}^{h-2k} \cdot \tilde{J}_{h-2k}=
\sum_{n=0}^{[{h\over 2}]-k} \rho_n(d-2,h-2k) {(h-2k)! \over 2^n n!
(h-2k-2n)!} \tilde{J}^{[n]}_{h-2k}\cdot \tilde{J}^{[n]}_{h-2k}
\end{equation}
including the propagator normalization (\ref{Q}) and using the
expressions of ${\tilde J}_{h-2k}$ from (\ref{tJ}) we can write the
current exchange for the ${\cal W}_q^s$ field (after a shift
$k+n=u$)
\begin{eqnarray}\label{A0}
{\cal A}(s,h)&=& (h!)^2 \sum_{k=0}^h {(s-h+2k)!! \over s!}
{{\tilde \rho}_k(d,h)^2 \over 2^{2k} (k!)^2 {\tilde \rho}_{k +
[{s-h \over 2}]}(d,s)} \cdot \\ \nonumber
 &&\sum_{u=k}^{[{h\over 2}]} {\rho_{u-k}(d-2,h-2k)
\over 2^{u-k} (u-k)! (h-2u)!} J^{[u]}_h \cdot J^{[u]}_h.
\end{eqnarray}
A change in the order of the summations leads to the expression
\begin{equation}\label{A}
{\cal A}(s,h)= {h!^2 \over s!}\sum_{u=0}^{[{h\over 2}]} {J^{[u]}_h
\cdot J^{[u]}_h \over 2^u (h-2u)!} \sum_{k=0}^{u} {(s-h+2k)!!
\over 2^k (k!)^2 (u-k)!} {{\tilde \rho}_k(d,h)^2 \over {\tilde
\rho}_{k + [{s-h \over 2}]}(d,s)}  \rho_{u-k}(d-2,h-2k).
\end{equation}
This is our final expression for the current exchanges written in
terms of irreducible fields. This expression has some remarkable
properties. In particular, taking $h=s$ and using (\ref{delta}) we get the
extremely simple expression
\begin{equation}\label{Ass}
A(s,s)= J_s \cdot J_s.
\end{equation}
We see that all traces of the currents have cancelled. We shall see
in the next section that this is exactly the result
which is obtained after using
 the propagator of the ${\cal W}_s^0=\varphi_s$ field.

What is even more remarkable is the fact that in the above result
the only non-vanishing contributions are those where $u \leq
max([{s-h \over 2}],[{h\over 2}])$. We have checked this property
numerically with the use of Mathematica for several values up to
$s=100$. This is perfectly consistent with the value one can expect for the
form of the propagator for the p-th trace of $\varphi_s$ (see
(\ref{proptra})), which gives further support to our results.

\setcounter{equation}0\section{Current Exchanges For
Triplets}\label{CET} In this section we will repeat the
computation performed in the previous Section using a different method. We will gauge fix the
Lagrangian in (\ref{Lcomp}) in a specific gauge in which the
fields $\varphi, D$ decouple from each other and we shall write
down the propagator for those fields in this gauge. Computing the current-current interaction we
will confirm that our result (\ref{A}) has the correct form and
the remarkable constrain on the current traces mentioned after (\ref{Ass})
appears naturally in perturbation theory formulated in terms of
reducible fields. We will also demonstrate the equivalence of the
two methods with several non-trivial examples.

\subsection{Gauge Fixing And The Propagator}\label{GF}

The most straightforward gauge is the one where
the auxiliary field $C_{s-1}$ in (\ref{Lcomp}) is set equal to zero.
The gauge fixing
($R_\xi$-gauge) term in the  Lagrangian has the form
\begin{equation}\label{Rxi}
{\cal L}_\xi= -{1\over 2\xi} C^2.
\end{equation}
The gauge fixing procedure  requires the
introduction of  a Faddeev-Popov determinant in the path integral
\begin{equation}\label{Z}
Z= \int d\omega e^{-{\omega^2 \over 2\xi}}\int [d\varphi] [dC]
[dD] \Delta_{FP}\delta(C-\omega)e^{{\cal L}+{\cal L}_\xi}
\end{equation}
where using the gauge transformation of $C$ from (\ref{deltaphi}) one gets
\begin{equation}\label{DFP}
\Delta_{FP}=det ({\delta C \over \delta \Lambda})=det (\Box).
\end{equation}
The FP determinant is field independent and can be absorbed into
the normalization constant of the path integral. Obviously, the
presence of interactions in the Lagrangian will make the FP
determinant field dependent, requiring therefore that we introduce
ghosts just as in QCD. In our case though the interaction
(\ref{ILW}) is abelian since it gauges the abelian rigid symmetries
of the free scalar Lagrangian (see i.e., \cite{Fotopoulos:2007yq}
and \cite{Bekaert:2009ud}) and ghost fields will not be needed. On the
contrary, when we consider the full interacting Lagrangian as i.e.
(\ref{KOmod})-(\ref{cyclic}), we will need FP ghost fields for a
consistent quantum theory. This, however, will not affect tree level
amplitudes in full analogy with QCD, since external states are
always on-shell, but it will play an important role in loop amplitudes
and in the optical theorem. We leave this and other interesting
issues for a future work, where it would be very interesting to
consider a consistent interacting theory beyond the tree level.

 The value of the parameter $\xi$
interpolates from Dedonder-Feynman gauge for $\xi=0$ to
Dedonder-Landau for $\xi=1$\footnote{ Notice that in the
literature for QED the $R_\xi$ gauge fixing Lagrangian is given by
${\cal L}_\xi= -{1\over 2\xi} (\partial_\mu A^\mu)$ and the
Feynman gauge corresponds to $\xi=1$. This corresponds to adding
(\ref{Rxi}) to (\ref{Lcomp}) for the spin-$1$ triplet, integrating
out $C$ first, which gives $C=\partial_\mu A^\mu$, and then gauge
fixing with a $\delta(\partial \cdot A -\omega)$ condition in the
path integral. This procedure gives us for $\xi=1$ the Feynman
gauge and for $\xi=0$ the Landau one. This is exactly the opposite
identification from the main text where gauge fixing $C$ as in
(\ref{Z}), rather than integrating it out, we get Landau gauge for
$\xi=1$ and Feynman gauge for $\xi=0$.} . The most useful gauge
for our purposes is the Feynman gauge which basically decouples
the field $C$ completely from the path integral and the gauge
fixed Lagrangian takes the form
\begin{equation}\label{Lgf}
{\cal L}+{\cal L}_{\xi=0}=- \, \frac{1}{2}\ (\partial_\mu
\varphi)^2 \ +  \ \frac{s(s-1)}{2} \ (\partial_\mu D)^2.
\end{equation}
It is  instructive to use the equations of motion for $C$ from
(\ref{eqm1}) to deduce the form of the gauge fixing in terms of
irreducible fields. Inserting (\ref{fD}) in the second equation of
(\ref{eqm1}) we get
\begin{equation}\label{gfPsi}
C= \partial\cdot \varphi- \partial D= \sum_{k=0}^{[{s\over 2}]-1}
{\tilde \rho}_k(d,s) \eta^k ( \partial \cdot \Psi_{s-2k}- {1\over
2} \partial \Psi'_{s-2k})
\end{equation}
where in the RHS we recognize immediately the Dedonder gauge
fixing term in the parentheses. The Feynman gauge fixing
corresponds to setting $C=0$ in the Lagrangian.

We  notice from (\ref{Lgf}) that the two fields $\varphi, D$ have
decoupled completely from each other, allowing us to simply invert
 their kinetic operators in order to get their
propagators
\begin{eqnarray}\label{propD}
\Delta(\varphi;\mu,\varphi;\nu)&=& {\eta_{\mu_1 (\nu_1}
\eta_{\mu_2
\nu_2}\dots \eta_{\mu_s \nu_s)}\over p^2 s!} \\
\Delta(D;\mu,D;\nu)&=& -{1\over s(s-1)}{\eta_{\mu_1 (\nu_1}
\eta_{\mu_2 \nu_2}\dots \eta_{\mu_{s-2} \nu_{s-2})}\over
p^2 (s-2)!} \nonumber \\
\Delta(\varphi;\mu, D)=0 \nonumber
\end{eqnarray}
where the parentheses in subscripts signify symmetrization with
respect to one set of the indices i.e., $\nu_1, \dots \nu_s$.
Notice the negative sign in the propagator of $D$. The field $D$
is a ``ghost" field. For other gauges like the Landau one, the
fields $\varphi, D$ do not decouple after integrating out the
auxiliary field $C$ and it is quite non-trivial to diagonalize the
kinetic operator to get the propagator of these states.

Let us see how this procedure works explicitly for the spin 2
case. The propagators for the irreducible fields $\Psi_2, \Psi_0$
from (\ref{Ls2}) in the Feynman gauge are
\begin{eqnarray}\label{s2fprop}
\Delta(\Psi_2;\mu, \Psi_2;\nu) &=& {\eta_{\mu_1\nu_1}\eta_{\mu_2
\nu_2}+ \eta_{\mu_1\nu_2}\eta_{\mu_2 \nu_1}- {2\over
d-2}\eta_{\mu_1\mu_2}\eta_{\nu_1 \nu_2} \over 2 p^2}
\nonumber \\
\Delta(\Psi_0, \Psi_0) &=& {d-2 \over p^2}
\end{eqnarray}
where we have taken into account the normalization factors of the
kinetic term of $\Psi_0$. Now using equations (\ref{Ws2}) we can
deduce the propagators for the fields $\varphi=\Psi_2+ {1\over
d+2} \eta \Psi_0 $ and $D= {1\over 2}\Psi'_2 + {1\over d-2}
\Psi_0$
\begin{eqnarray}\label{s2wfprop}
\Delta(\varphi;\mu, \varphi;\nu) &=& {\eta_{\mu_1\nu_1}\eta_{\mu_2
\nu_2}+
\eta_{\mu_1\nu_2}\eta_{\mu_2 \nu_1} \over 2 p^2} \\
\Delta(D,D)&=& -{1\over 2 p^2} \\
\Delta(\varphi;\mu, D)=0 \nonumber
\end{eqnarray}
where we easily see that the fields $\varphi, D$ are decoupled
from each other and the propagators agree with (\ref{propD}). The
fact that this decoupling is special to the Feynman gauge can be
seen if we repeat the procedure described above in Landau gauge where the
propagators take the form
\begin{eqnarray}\label{s2lprop}
\Delta(\Psi_2;\mu, \Psi_2;\nu) &=& \frac{\eta_{\mu_1\nu_1}\eta_{\mu_2
\nu_2}+ \eta_{\mu_1\nu_2}\eta_{\mu_2 \nu_1}}{2p^2} \\ \nonumber
&&-\frac{\frac{2\eta_{\mu_1\mu_2}\eta_{\nu_1 \nu_2}}{d-2}+
\frac{ \eta_{\mu_1\nu_1}
p_{\mu_2} p_{\nu_2} \ + \ 3 \ permutations}{p^2}}{2p^2}
\nonumber \\
\Delta(\Psi_0, \Psi_0) &=& {d-2 \over p^2}.
\end{eqnarray}
From the form of the propagators above we can easily compute the
propagator  $\Delta(\varphi;\mu, D)$ and indeed we find that it is
non-zero and therefore the fields do not decouple.

\subsection{Current Exchange and Comparison}\label{CEC}

In this section we will compute the current-current interaction
for the Lagrangian (\ref{LWint}) using the propagators in
(\ref{propD}). For this we will need to compute the propagators
for arbitrary traces of the fields
$\Delta(\varphi^{[p]},\varphi^{[p]})$. There are obviously
propagators of the form $\Delta(\varphi^{[p]},\varphi^{[q]})$ for
$p \neq q$ but they will not be needed for the current exchange
computations we consider in this note. From the form of the
propagators in (\ref{propD}) and the use of (\ref{etak}) we can
deduce the general form of these propagators
\begin{equation}\label{proptra}
\Delta(\varphi_s^{[p]}; 1,\varphi_s^{[p]}; 2)=
\sum_{k=max(2p-[{s\over 2}])}^p B_k(s,p) \eta_1^{p-k} \eta_2^{p-k}
\Delta(\varphi_{s-4p+2k}^{[p]}; 1,\varphi_{s-4p+2k}^{[p]}; 2)
\end{equation}
where $p \in [0,[{s\over 2}]]$.  The notation needs some
explanation. The two sets of indices for the two fields of the
propagators are denoted in a shorthand notation in which, $1$
stands for the $\mu_1, \mu_2, \dots \mu_{s-2p}$ subscripts of the
first field and $2$ for the $\nu_1, \nu_2, \dots \nu_{s-2p}$
subscripts of the second field. The propagators in the summation
on the RHS are the usual propagators of (\ref{propD}) for the
field with spin $s-4p-2k$. The coefficients $B_k(s,p)$ are
unknowns to be determined by the explicit calculation. This has
not been achieved for the moment in the general case. The lower
bound in the k-summation is obvious since the minimum spin of the
propagators in the RHS is zero.

Now, an important observation is that each
$\eta^{p-k}_1\eta^{p-k}_2$ term will give us a current-current
interaction proportional to $J^{p-k}_s \cdot J^{p-k}_s$. The lower
bound of the summation means that the maximum trace of the
currents allowed in a $\varphi^{[p]}_s$ exchange can be written as
\begin{equation}\label{constr}
max([{s\over 2}]-p,p)=max( {s-h\over 2},[{h\over 2}])
\end{equation}
We immediately recognize the bound we deduced numerically from (\ref{A}).
 The equation
 (\ref{constr})  explains the surprising constrain which we pointed
out in the last paragraph of the Section \ref{Decomp}. This fact
strongly indicates that the procedures described in Section
\ref{Decomp} and Section \ref{CET} are completely equivalent.
 This was not obvious at all in the expression of
(\ref{A}) but it is a direct consequence of our
$\Delta(\varphi_s^{[p]}; 1,\varphi_s^{[p]}; 2)$ propagators in the
theory formulated in terms of
$\varphi$ and  $D$.

Since we have not achieved to the moment to compute the explicit
expressions of the $B_k(s,p)$ coefficients we will proceed with a
few examples which establish the equivalence of the two methods we
have used. Let us define
\begin{equation}\label{C}
C(s-h,u,h) \equiv \sum_{k=0}^{u} {(s-h+2k)!! \over 2^k k!^2
(u-k)!} {{\tilde \rho}_k(d,h)^2 \over {\tilde \rho}_{k + [{s-h
\over 2}]}(d,s)}  \rho_{u-k}(d-2,h-2k)
\end{equation}
which appears in (\ref{A}). Numerical computation of the
coefficients $C(s-h,u,h)$ gives zero for $u > {s-h \over 2}$.

\paragraph {Spins 2 and 4} We can easily compute the relevant
coefficients $C(s-h,u,h)$ we will need for the spin 2 and 4 cases
\begin{eqnarray}\label{C4}
C(2,0,h) =2 (d+2(h-1)) & \qquad & C(2,1,h) =2 \nonumber \\
C(4,0,h) =8 (d+2h-2)(d+2h) & \qquad & C(4,1,h) =16 (d+2h-2) \nonumber \\
& C(4,2,h)= 4. &
\end{eqnarray}

From the explicit expression of the propagators in (\ref{propD}),
the interaction Lagrangian (\ref{ILW}) and the definition of
${\cal W}_s^q$ in (\ref{hatS}) we get for the corresponding
current exchanges
\begin{eqnarray}\label{s2CE}
{\cal A}(2,2) &=& p^2 J_2\Delta(W^0_2,W^0_2) J_2=  J_2\cdot J_2\\
{\cal A}(2,0) &=& p^2 J_0\Delta(W^1_2,W^1_2) J_0 = (d-2)J_0\cdot
J_0\nonumber
\end{eqnarray}
where ${\cal A}(s,h)$ is the current exchange for the ${\cal
W}_q^s$ field as in (\ref{A}). The current exchange is defined as
the residue of the corresponding Feynman diagram for the
current-current interaction. We can compare (\ref{s2CE}) now with
the expression from (\ref{A}) using (\ref{C4}). We verify that
${\cal A}(2,0) = {C(2,0,0)\over 2} J_0\cdot J_0$ while ${\cal
A}(2,2)$ is given by (\ref{Ass}) as expected.

The spin 4 case requires that we compute the propagators of traces
of fields. The relevant propagators, after a short computation,
are given by
\begin{eqnarray}\label{props4}
\Delta(\varphi';\mu,\varphi';\nu) &=& 2 { (d+4) \eta_{\mu_1
(\nu_1} \eta_{\mu_2 \nu_2)} + 2\eta_{\mu_1 \mu_2}
\eta_{\nu_1\nu_2} \over
4! p^2} \nonumber \\
\Delta(\varphi^{''},\varphi^{''})&=& 8{ d(d+2) \over 4! p^2}  \nonumber\\
\Delta(D;\mu,D;\nu)&=& -{1\over 12}{  \eta_{\mu_1 (\nu_1}
\eta_{\mu_2 \nu_2)}\over 2! p^2} \\
\Delta(D',D')&=& -{d\over 12 p^2}.\nonumber
\end{eqnarray}
For $h=2$ we get
\begin{equation}
{\cal A}(4,2)= p^2 J_2 \Delta({\cal W}_4^1,{\cal W}_4^1) J_2=
{2(d+2) J_2\cdot J_2 + 2 J_{2}^{[1]}\cdot J_{2}^{[1]}\over 12}
\end{equation}
which is reproduced by the expression in (\ref{A})
\begin{equation}
{\cal A}(4,2)= {(2!)^2 \over 4!} \sum_{u=0}^1 {C(2,u,2) \over
(2-2u)! 2^u}J_{2}^{[u]}\cdot J_{2}^{[u]}.
\end{equation}
In a similar manner we obtain that ${\cal A}(4,0)= {8 d(d-2) \over
12}J_0\cdot J_0$. So we have shown that both methods agree for
spin 2 and spin 4.

\paragraph {Spin s and h=s-2 case}
~For ~this ~case ~we ~need ~to ~compute ~the ~propagator
$\Delta(\varphi';1,\varphi';2)$. ~Keeping ~a ~more ~compact ~notation
we find
\begin{equation}\label{propf'}
\Delta(\varphi',\varphi')= {2 \over p^2} ( {d+2(s-2) \over s(s-1)}
\Delta_{s-2}(1,2) + 2 {\eta _1 \eta_2 \Delta_{s-4}(1,2) \over
s(s-1)(s-2) s-3)} ).
\end{equation}
The subscripts of $\Delta$ on the RHS are the spin of the
propagator from (\ref{propD}) and we have suppressed the field
variables and their space-time indices. The second term on the RHS
implies symmetrization of each $\eta_i,i=1,2$ with the
corresponding indices of the $\Delta_{s-4}(1,2)$ propagator. When
the propagator is contracted with symmetric currents $J_{s-2}$ for
both sets of indices then there are ${(s-2)(s-3)\over 2}$ terms
from the symmetrization of each $\eta_i$ which result into
${(s-2)(s-3)\over 2}$ traces of the currents $J^{[1]}_{s-2}$ for
each set of indices. The final result we get is
\begin{eqnarray}
{\cal A}(s,s-2)&=& p^2 J_{s-2} \Delta({\cal W}_s^1,{\cal W}_s^1)
J_{s-2} \\ \nonumber &&=2  {d+2(s-3) \over s(s-1)} J_{s-2} \cdot
J_{s-2} + {(s-2)(s-3) \over s(s-1)} J^{[1]}_{s-2}\cdot
J^{[1]}_{s-2}.
\end{eqnarray}
The same computation should be reproduced by (\ref{A})
\begin{equation}
{\cal A}(s,s-2)= {(s-2)!^2 \over s!} \sum_{u=0}^{[{s\over 2}]-1}
{C(2,u,s-2) \over (s-2-2u)! 2^u}J^{[u]}_{s-2}\cdot J^{[u]}_{s-2}.
\end{equation}
Taking into account the constraint (\ref{constr}) we see that only
$u=0,1$ survive and a trivial computation using (\ref{C}) gives
agreement of the two results.

Further examples become more and more tedious.
Nevertheless, we believe that we have demonstrated in a sufficient manner the
equivalence of the two methods. Actually our result in (\ref{A})
can be used to extract the $B_k(s,p)$ coefficients rather than
trying to compute them directly by taking traces of (\ref{propD}).

\section{Conclusions}\label{Con}
In this paper we developed a technique of constructing propagators for massless
irreducible
higher spin modes from the Lagrangians describing the reducible higher spin fields.
The main motivation for this is that often the gauge-invariant Lagrangians describing the
reducible higher spin modes have a much simpler form than those describing
irreducible higher spin modes, especially when one considers interacting theories.
This technique can be straightforwardly generalized to the case of massive
higher spin fields. As an application we considered the current-current exchange
amplitudes
obtained from the cubic Lagrangians describing an interaction of higher spin fields with scalars.

It would be interesting to generalise these results for the case of $AdS$
space and for the triplets containing fermionic fields.
Another interesting application of our  results  may be
a computation of the higher order scattering amplitudes for systems which contain
exact vertexes  in all orders in the coupling constant. A possible example of such kind of systems
is given in the present paper for massive and in
 \cite{Fotopoulos:2007nm} for massless bosonic higher spin fields.

\section{Acknowledgements}
We are grateful to C.Angelantonj and A. Sagnotti for useful
discussions. We would also like to thank the referee for several
interesting points we have clarified and correcting erroneous
factors in a formula. The work of A. F. was supported by an INFN
postdoctoral fellowship`and partly supported by the Italian
MIUR-PRIN contract 20075ATT78. The work of M.T. was supported by a
STFC rolling grant ST/G00062X/1.

\renewcommand{\thesection}{A}
\setcounter{equation}{0}
\renewcommand{\theequation}{A.\arabic{equation}}
\section{Appendix A: Decoupling of the Lagrangian for Spin 2 and Spin 4
triplet}\label{AppA}

The Lagrangian for spin s=4 in (\ref{Lsym}) is given by
\begin{eqnarray}\label{Ls4}
{\cal L} &=& -{1\over 10} (\partial_{(k} \varphi_{\mu \nu \rho
\sigma)})^2 +4 ((\partial \cdot \varphi _{\nu \rho \sigma}))^2 -4
(\partial \cdot \varphi_{\nu \rho \sigma})\ D^{\nu\rho\sigma}
\nonumber \\
&& +4 (\partial_{(k} D_{\mu \nu )})^2 -12 ((\partial \cdot
D_\mu))^2
\end{eqnarray}
We insert the decomposition (\ref{phiDs4}) and using (\ref{etak})
we get from each of the five terms of the Lagrangian (\ref{Ls4}) the cross-terms
listed in Table 1.
\begin{equation}\label{table}
\begin{array}{|c|c|c|c|c|c|c|}
   \hline
   M.T&-{1\over 10} (\partial \varphi)^2& +4 (\partial \cdot
   \varphi)^2&-4(\partial \cdot \varphi ) D&+4 (\partial D)^2&-12 (\partial \cdot
D)^2& Total \\
\hline {(\partial \Psi'_4) (\partial \Psi_2) \over d+2} & -2 &
0&-2&4&0&0 \\
\hline {(\partial \cdot \Psi_4) (\partial \Psi_2) \over
d+2}&-4&8&-4&0&0&0 \\
\hline { (\partial \cdot \Psi'_4) (\partial \cdot \Psi_2)
\over d+2}&0&24&-12&0&-12&0 \\
\hline { (\partial \cdot \Psi'_4) (\partial \Psi'_2) \over
d+2}&0&0&-6&12&-6&0 \\
\hline{(\partial \cdot \Psi'_4) (\partial \Psi_0) \over
d(d-2)}&-12&24&-24&24&-12&0 \\
\hline {(\partial \cdot \Psi_2)(\partial \Psi_0)\over
d(d+2)(d-2)}&
-12(d+4)&24(d+4)&-12(d+6)&+48&-24&0 \\
\hline {(\partial \Psi'_2)(\partial\Psi_0)\over
d(d+2)(d-2)}&-6(d+4)&24&-6(d+6)&12(d+4)&-12&0 \\
\hline
\end{array} \nonumber
\end{equation}
\begin{center}
Table \ 1
\end{center}
In table 1 the numbers in each column indicate the coefficient of
the cross terms of $\Psi_n$ (first column) from each term of the
Lagrangian (\ref{Ls4}) (first row). ``M.T'' stands for ``Mixing Terms''.

As we can see all mixing terms have vanishing coefficients and we
confirm that the Lagrangian decomposes to gauge invariant
Lagrangians for $\Psi_4,\Psi_2,\Psi_0$. Since the fields $\Psi_n$
are by definition doubly traceless and transform with traceless
parameters (\ref{tLambda}),  we can deduce easily that the
Lagrangians for the irreducible fields are Fronsdal Lagrangians.
We are only missing the correct normalization coefficient for each
one of them. This will be computed in  Appendix B for the general
spin-s case. A few examples of our calculations from the first
term in the Lagrangian are
\begin{eqnarray}\label{ex}
(\partial \Psi_4)(\eta \partial \Psi_2)&=& 10[ \partial \Psi'_4 +2
(\partial \cdot \Psi_4)] (\partial \Psi_2) \nonumber \\
(\eta \partial \Psi_2)^2&=& 10 [(d+6) (\partial \Psi_2)^2 +12
(\partial \cdot \Psi_2)^2 + 3 (\partial \Psi'_2)^2 +12 (\partial
\cdot \Psi_2)(\partial \Psi'_2)]\nonumber \\
(\partial \Psi_4)(\eta^2 \partial \Psi_0)&=&60(\partial \cdot
\Psi'_4) (\partial \Psi_0) \\
(\eta \partial \Psi_2)(\eta^2 \partial \Psi_0)&=& 30(d+4)
(2(\partial \cdot \Psi_2) + \partial \Psi'_2) (\partial \Psi_0)
\nonumber
\end{eqnarray}
Let us explain a bit how we get the result of the first equation
in (\ref{ex}). The expression on the LHS has total 10 terms. These
come from the symmetrization of $\eta$ with $\partial \Psi_2$.
Then depending which terms $\eta$ contracts from $\partial \Psi_4$
we get $\eta_{\mu \nu}
\partial^\mu \Psi_4^{\nu \rho \sigma \lambda}$ or $\eta_{\mu \nu}
\partial^\rho \Psi_4^{\mu \nu \sigma \lambda}$. In this way we obtain the two terms on the
RHS with the given multiplicities.

\renewcommand{\thesection}{B}
\setcounter{equation}{0}
\renewcommand{\theequation}{B.\arabic{equation}}
\section{Appendix B: Lagrangian Normalization for Irreducible
Fields}\label{AppB} In this appendix we will prove equation
(\ref{Q}) of the main text.

Using (\ref{etak}) and various manipulations we have the following
identities
\begin{eqnarray}\label{ident}
(\eta^k \partial \Psi_{s-2k})^2 &=& {(s+1)! \over {\tilde
\rho}_k(d,s+2) 2^k k! (s-2k+1)!} (\partial \Psi_{s-2k})^2 + \dots
\nonumber \\
(\partial \cdot (\eta^k  \Psi_{s-2k}))^2 &=& {(s-1)! \over {\tilde
\rho}_{k-1}(d,s) 2^{k-1} (k-1)! (s-2k+1)!} (\partial
\Psi_{s-2k})^2
+ \dots \\
(\eta^{k-2} \partial \Psi_{s-2k})^2 &=& {(s-3)! \over {\tilde
\rho}_{k-2}(d,s-2) 2^{k-2} (k-2)! (s-2k+1)!} (\partial
\Psi_{s-2k})^2 + \dots \nonumber
\end{eqnarray}
where the dots are all terms involving traces and divergences of
the fields $\Psi_{s-2k}$ and we have used the following
combinatorial identity for symmetrized tensors contracted with
symmetrized tensors
\begin{equation}\label{combi}
\eta^k T_q \rightarrow \eta^k \times T_q {(2k+q)! \over (2k)! q!}.
\end{equation}.
Notice that (\ref{combi}) is not an equality. What it means is
that if the tensor of the LHS is contracted with a tensor totally
symmetric over all the indices then we can substitute it with the
expression on the RHS. The symbol $\times$ means the product of
the two tensors without symmetrization of their indices. We also
remind the reader that as in \cite{Francia:2002pt} the notation
$\eta^k$ has a total of $(2k-1)!!$ terms unlike the symmetrization
of a $2k$-rank tensor which has $(2k)!$ terms.

Using the identities above we can insert (\ref{fD}) in
(\ref{Lsym}) dropping the cross-terms and keeping only the
 $(\partial \Psi_{s-2k})^2$ terms which are relevant to our case. We get
\begin{eqnarray}\label{norm1}
&&{s!\over 2^{k+1} k! (s-2k+1)!} {{\tilde \rho}^2_{k}(d,s)\over
{\tilde \rho}_{k-1}(d,s)} \cdot \\ \nonumber
&& \left( d-2(s-k) + {4(k-1)k \over
d+2(s-k-1)} -4k \right)(\partial \Psi_{s-2k})^2 + \dots
\end{eqnarray}
where we have used the identities
\begin{eqnarray}\label{rhoiden}
{\tilde \rho}_{k}(d,s+2)&=&{{\tilde \rho}_{k-1}(d,s)\over
d+2(s-k)} \\
{\tilde \rho}_{k-1}(d,s)&=&{{\tilde \rho}_{k-2}(d,s-2)\over
d+2(s-k-1)}. \nonumber
\end{eqnarray}
Now we can use the following identity (which was used also to
convert (\ref{Lcomp2}) to (\ref{Lsym}))
\begin{equation}\label{idensym}
(\partial_\mu \Psi_q)^2={1\over q+1} (\partial \Psi)^2 - q
(\partial \cdot \Psi_q)^2
\end{equation}
to arrive finally at
\begin{equation}\label{norm2}
{\cal L}_{\varphi, D} \rightarrow -{1\over 2}{\tilde
\rho}_k(d,s)^2 {s! \over 2^k k! (s-2k)!}(\partial_\mu
\Psi_{s-2k})^2
\end{equation}
after we have used the identity
\begin{equation}\label{rhoiden2}
{\tilde \rho}_{k}(d,s)= {\tilde \rho}_{k-1}(d,s)
{(d+2(s-2k))(d+2(s-2k-1)\over d+2(s-k-1))}.
\end{equation}
(\ref{norm2}) should be contrasted to the coefficient of the first
term in (\ref{Fronsdal}).

\end{document}